
\input phyzzx
\hoffset=0.375in
\overfullrule=0pt
\def\kms{\rm km\,s^{-1}}

\font\bigfont=cmr17
\centerline{}
\bigskip
\bigskip
\singlespace
\centerline{\bigfont Self-Lensing By A Stellar Disk}
\bigskip
\centerline{\bf Andrew Gould}
\bigskip
\centerline{Dept of Astronomy, Ohio State University, Columbus, OH 43210}
\smallskip
\centerline{E-mail:  gould@payne.mps.ohio-state.edu}
\bigskip
\bigskip
\vskip 1.5in
\singlespace
\centerline{\bf ABSTRACT}

	I derive a general expression for the optical depth $\tau$
for gravitational lensing of stars in a disk by Massive Compact Objects
(Machos) in the same disk.  For the more restricted case where the disk
is self-gravitating and the stars and Machos have
the same distribution function,  I find $\tau = 2\VEV{v^2}/c^2\sec^2 i$
where $\VEV{v^2}$ is the mass-weighted vertical velocity dispersion, and
$i$ is the angle of inclination.
This result does not depend on any assumptions about the velocity distribution.
As an example, if stars within the bar of the Large Magellanic Cloud (LMC)
account for the observed optical depth $\tau\sim 8\times 10^{-8}$ as has
recently been suggested, then $v\gsim 60\,\kms$.  This is substantially
larger than the measured dispersions of known LMC populations.

Subject Headings:  dark matter -- gravitational lensing
\vskip 0.3in
\centerline{submitted to {\it The Astrophysical Journal}: August 1, 1994}
\bigskip
\centerline{Preprint: OSU-TA-13/94}
\endpage
\normalspace

\chapter{Introduction}

	Alcock et al.\ (1993) and Aubourg et al.\ (1993) have recently
reported the detection of several candidate microlensing events toward
the Large Magellanic Cloud (LMC).  These discoveries have stimulated
a number of authors to consider the lensing of the LMC stars by Massive
Compact Objects (Machos) in the bar of the LMC itself (Wu 1994; Sahu 1994).
Here I present some very general results on self-lensing by a distant
galactic disk and apply these results to the LMC.  The results can also
be applied to lensing by the disk of M31 (e.g.\ Gould 1994b), and with
some modification to self-lensing by the bulge of the Milky Way.

\chapter{Self-Lensing By A Disk}

	Consider a galactic disk at a distance $d$ that
is large compared to the disk thickness.  The optical depth to lensing
of a disk star at distance $d+z$ is
$$\tau(z) = {4 \pi G \over c^2}\int_{-\infty}^z d y\,\rho_m(y\cos i){(z-y)(d+y)
\over d+z}\rightarrow{4 \pi G \over c^2}\int_{-\infty}^z d y\,\rho_m(y\cos i)
(z-y),\eqn\tauone$$
where $i$ is the angle of inclination and $\rho_m(z)$ is the mass density of
Machos as a function of height above the plane
The mean observed optical depth over the population of all stars in the
disk is
$$\tau = {4 \pi G \over c^2}N_s^{-1}\int_{-\infty}^\infty d z\, n_s(z)
\int_{-\infty}^z d y\,\rho_m(y\cos i)(z-y),\eqn\tautwo$$
where $n_s$ is the number density of sources and
where I define
$$N_s \equiv \int_{-\infty}^\infty d z\, n_s(z),\qquad
\Sigma_m \equiv \int_{-\infty}^\infty d z\, \rho_m(z).\eqn\ndef$$
Equation \tautwo\ can be evaluated,
$$\tau = {\pi G \Sigma_m\over c^2}\sec^2 i\int_{-\infty}^\infty d z\,
[1 - F_m(z)][1 + F_s(z)],\eqn\tauthree$$
where
$$F_m(z) \equiv -1+{2\over \Sigma_m}\int_{-\infty}^z d y\, \rho_m(y),\quad
F_s(z) \equiv -1+{2\over N_s}\int_{-\infty}^z d y \,n_s(y).\eqn\cumdist$$
For the most part, I will restrict attention to the case where $\rho$ and $n$
are symmetric in $z$, in which case
$$F_m(z) \equiv {2\over \Sigma_m}\int_0^z d y \rho_m(y),\eqn\cumdisttwo$$
and similarly for $F_n(z)$, so that
$$\tau = {2\pi G \Sigma_m\over c^2}\sec^2 i\int_0^\infty d z\,
[1 - F_m(z)][1 + F_s(z)].\eqn\taufour$$

	For the case where the stars and Machos have the same distribution,
equation \taufour\ becomes
$$\tau = {2\pi G \Sigma_m\over c^2}\sec^2 i\int_0^\infty d z\,
\{1 - [F_m(z)]^2\}\quad ({\rm pure}\ {\rm self-lensing}).
\eqn\taufive$$
On the other hand, suppose that the stars are confined to a thin planar
distribution of much smaller scale height than the Machos.  In this case
$F_n(z)=\Theta(z)$ where $\Theta$ is the step function, so that
$$\tau = {2\pi G \Sigma_m\over c^2}\sec^2 i\int_0^\infty d z\,
[1 - F_m(z)]\quad ({\rm thin}\ {\rm stellar}\ {\rm disk}).
\eqn\tausix$$
Note that equation \tausix\ is always smaller than equation \taufive.

\chapter{Self-Gravitating Disk}

	Suppose that the source number distribution is proportional to the
Macho mass distribution and let the disk potential, $\Psi(z)$ be due entirely
to the Machos.
Then the Jeans equation,
$${d [\rho_m(z) \overline{v^2(z)}]\over d z} = - \rho_m(z){d\Psi\over d z},
\eqn\Jeans$$
can be used to evaluate the optical depth \taufive.  In this
case, $d\Psi/d z = 2\pi G\Sigma_m F_m(z)$, so that
$$\rho_m(z)\overline{v^2(z)} = {\pi G \Sigma_m^2}\int^z d y\, F(y)
{d F \over d y} = {\pi G \Sigma_m^2\over 2}\{1- [F(z)]^2\},\eqn\press$$
where I have evaluated the integration constant using the fact that the
pressure vanishes at infinity.  Substituting equation \press\ into equation
\taufive, I find
$$\tau = 2{\VEV{v^2}\over c^2}\sec^2 i,\eqn\final$$
where $\VEV{v^2}$ is the mass weighted velocity dispersion,
$$\VEV{v^2} \equiv {2\over \Sigma_m}\int_0^\infty d z\,
\rho_m(z)\overline{v^2}(z).\eqn\mweight$$

\chapter{Some Additional Effects}

	Even if the stars and Machos have the same distribution, there are
two effects which tend to reduce the optical depth relative to equation
\final.  First, the disk will in general have some mass (e.g., a thin sheet
of gas) that provides neither sources nor lenses.  Any such non-lensing
mass distribution will increase $|F_m(z)|$ for all values of $z$ and hence
will, by equation \taufive\ reduce the optical depth.  The effect is not
large however.  In the most extreme case where there is a very large
amount of such material near the plane, an isothermal population of
stars would be distributed
as $\rho_m(z) = (\Sigma_m/2 h) \exp(-|z|/h)$, where $h$ is the scale height.
Then $F_m(z)=1-\exp(-|z|/h)$, so that $\tau = 3\pi G\Sigma_m h/c^2$.  That is,
$\tau = 1.5 \VEV{v^2}/c^2$, a 25\% reduction relative to the
pure self-gravitating disk.

	Second, there is extinction.  Here, the stars on
the far side of the disk (which have higher optical depths than the near
side) will be preferentially eliminated leading to lower optical depths.
In M31, where $i\sim 78^\circ$ so that
the dust at the midplane of the disk is probably opaque to optical light, the
effect can be very large (Gould 1994b).  From equation \tauthree, I find
that the optical depth is reduced by a factor of $\ln 2/2$ for an isothermal
self-gravitating disk [$F_m(z)=\tanh(z/D)$], and by 1/3 for an exponential
disk.  For the LMC, however, the effect is small.

\chapter{Nearly Face-on Galaxies}

As a practical matter, for nearly face-on galaxies one estimates the vertical
velocity dispersion from the line of sight
dispersion.  If the velocity ellipsoids of other galaxies have axis ratios
4:3:2 like the Milky Way, then the line of sight dispersion overestimates
the vertical dispersion, $\VEV{v_z^2}$ by a factor $\sim (1 + 2\sin^2 i)$.
Hence, for galaxies $i\lsim 45^\circ$, the upper limit
$\tau<2\VEV{v^2}/c^2\sec^2 i$ derived in the
previous two sections remains valid provided that one uses the line of sight
dispersion to estimate the vertical dispersion.

\chapter{Application to the LMC}

	The observed line of sight velocity dispersion of CH stars in
the inner parts of the LMC is $\sim 20\,\kms$ (Cowely \& Hartwick 1991).
If the disk LMC
stars have a similar dispersion, the upper limit to their contribution
to the optical depth is $\tau < 1\times 10^{-8}$.
I have previously
estimated (Gould 1994a) the observed optical depth to be
$\tau\sim 8\times 10^{-7}$ on the basis of the initial reports of
Alcock et al.\ (1993).  Subsequently, a larger area has been searched and
two more events have been found (K.\ Griest 1994, private communication),
but my estimate remains unchanged.  Therefore, unless the general population
of stars in the LMC have
a substantially larger dispersion than the CH stars, they cannot contribute
significantly to the observed optical depth.
The suggestion of Sahu (1994) that the observed optical depth is due to stars
in the LMC would require that the line-of-sight dispersion be $v_z\gsim 60\,
\kms$, far in excess of the dispersion of any known population.


\bigskip
\bigskip
\Ref\al{Alcock, C.\ et al.\ 1993, Nature, 365, 621}
\Ref\al{Aubourg, E.\ et al.\ 1993, Nature, 365, 623}
\Ref\Cow{Cowley, A.\ P., \& Hartwick, F.\ D.\ A.\ 1991, ApJ, 373, 80}
\Ref\gould{Gould, A.\ 1989, ApJ, 341, 748}
\Ref\goulda{Gould, A.\ 1994a, ApJ, 421, L71}
\Ref\gouldb{Gould, A.\ 1994b, ApJ, 435, in press}
\Ref\sahu{Sahu, K.\ C.\ 1994, Nature, in press}
\Ref\Wu{Wu, X.-P.\ 1994, ApJ, in press}
\refout
\end